\pdfoutput=1

\documentclass[iop]{emulateapj}

\usepackage{graphicx}
\usepackage{amsmath}
\usepackage{relsize} 
\usepackage{soul} 

\usepackage{color} 

\usepackage{amsmath,tikz} 
\usetikzlibrary{arrows,shapes,positioning}

\usepackage{listings} 

\usepackage{natbib}
\usepackage{aas_macros}
\usepackage{hyperref}

\def\danger{\fontencoding{U}\fontfamily{futs}\selectfont\char 66\relax}

\def\xd{\textsf{X3D}}
\def\html{\textsf{HTML}}
\def\pdf{\textsf{PDF}}
\def\ud{\textsf{U3D}}
\def\stl{\textsf{STL}}
\def\fits{\textsf{FITS}}
\def\txt{\textsf{TXT}}
\def\png{\textsf{PNG}}
\def\wrl{\textsf{VRML}}

\def\python{\textsc{python}}
\def\mayavi{\textsc{mayavi}}
\def\matplotlib{\textsc{matplotlib}}
\def\yt{\textsc{yt}}
\def\firefox{\textsc{firefox}}
\def\chrome{\textsc{chrome}}
\def\safari{\textsc{safari}}
\def\ie{\textsc{internet explorer}}
\def\blender{\textsc{blender}}
\def\pdfrg{\textsc{pdf3dreportgen}}
\def\ffmpeg{\textsc{ffmpeg}}
\def\meshlab{\textsc{meshlab}}
\def\xdom{\textsc{x3dom}}
\def\astroblend{\textsc{astroblend}}

\def\HI{\normalfont H\,{\smaller \textsc I}}

\shorttitle{The {\xd} pathway}
\shortauthors{Vogt et al.}

\begin{document}

\title{Advanced Data Visualization in Astrophysics: the {\xd} Pathway}

\author{Fr\'ed\'eric P.A. Vogt\altaffilmark{1}, Chris I. Owen\altaffilmark{1}, Lourdes Verdes-Montenegro\altaffilmark{2}, Sanchayeeta Borthakur\altaffilmark{3}}
\email{frederic.vogt@alumni.anu.edu.au}

\altaffiltext{1}{Research School of Astronomy and Astrophysics, Australian National University, Canberra, ACT 2611, Australia.}
\altaffiltext{2}{Instituto de Astrof\'iõsica de Andaluc\'ia, CSIC, Apdo. Correos 3004, E-18080 Granada, Spain.}
\altaffiltext{3}{Department of Physics and Astronomy, Johns Hopkins University, 3400 N. Charles Street, Baltimore, MD 21218, USA.}

\begin{abstract}
Most modern astrophysical datasets are multi-dimensional; a characteristic that can nowadays generally be conserved and exploited scientifically during the data reduction/simulation and analysis cascades. Yet, the same multi-dimensional datasets are systematically cropped, sliced and/or projected to printable two-dimensional (2-D) diagrams at the publication stage. In this article, we introduce the concept of the ``{\xd} pathway'' as a mean of simplifying and easing the access to data visualization and publication via three-dimensional (3-D) diagrams. The {\xd} pathway exploits the facts that 1) the {\xd} 3-D file format lies at the center of a product tree that includes interactive {\html} documents, 3-D printing, and high-end animations, and 2) all high-impact-factor \& peer-reviewed journals in Astrophysics are now published (some exclusively) online. We argue that the {\xd} standard is an ideal vector for sharing multi-dimensional datasets, as it provides direct access to a range of different data visualization techniques, is fully-open source, and is a well defined ISO standard. Unlike other earlier propositions to publish multi-dimensional datasets via 3-D diagrams, the {\xd} pathway is not tied to specific software (prone to rapid and unexpected evolution), but instead compatible with a range of open-source software already in use by our community. The interactive {\html} branch of the {\xd} pathway is also actively supported by leading peer-reviewed journals in the field of Astrophysics. Finally, this article provides interested readers with a detailed set of practical astrophysical examples designed to act as a stepping stone towards the implementation of the {\xd} pathway for any other dataset.
\\
\end{abstract}

\keywords{general: publications, bibliography -- general: standards -- methods: data analysis -- techniques: miscellaneous}

\section{Introduction}\label{sec:intro}

A large majority of datasets in modern Astrophysics, whether observational or theoretical, are multi-dimensional with a number of dimensions $n>2$. The multi-dimensionality of a dataset can be intrinsic: for example the (reduced) three-dimensional (3-D) data cubes acquired by integral field spectrographs. Alternatively, the multi-dimensionality of a dataset can be the result of the combination of several subsets of data, for example when looking at the Fundamental Plane formed by elliptical galaxies \citep{Springbob12}. Understanding and characterizing the structure and content of these complex datasets across all their dimensions holds the key to improving our understanding of the physics at play in the Universe. 

The computing power nowadays accessible to astronomers usually allows multi-dimensional datasets to be handled as such throughout the entire data reduction/simulation and analysis cascades. At the result publication stage however, multi-dimensional datasets are systematically sliced, compressed and/or projected: a ``habit'' largely driven by historical reasons. As dedicated scientific journals were physically printed on papers, it was a \textit{physical requirement} for scientific diagrams (used as a mean of illustrating the content of complex datasets) to be printable. 

Forcing a multi-dimensional dataset on a two-dimensional (2-D) diagram usually implies a loss of information. Such a loss of information is certainly not always detrimental, as it can help emphasize certain elements of interest in the data. In a growing number of cases however, the need to understand the full multi-dimensional structure of the data - and our ability as scientists to share such structures with the community - is directly and evidently affected by the current lack of suitable alternatives to the classical 2-D \textit{printable diagrams}. 

The exploration of alternative techniques for data visualization is not only warranted by the existing difficulties associated with sharing multi-dimensional datasets in a clear, concise and meaningful way, but also by the data ``deluge'' triggered by the most recent or forthcoming astronomy facilities. For example, data products from the \textit{Low-Frequency Array for Radio Astronomy} (\textit{LOFAR}) or the \textit{Square Kilometre Array} (\textit{SKA}) by 2020 can not be easily handled with current computing power and technologies. Visualizing and sharing such complex (and large) datasets with the community will inevitably require innovative solutions.

All major scientific journals dedicated to astrophysical sciences are now published online. Some, like the \textit{American Astronomical Society journals}\footnote{The \textit{AAS journals} include (at the time of submission of this article) the \textit{Astrophysical Journal} (\textit{ApJ}), \textit{Astrophysical Journal Letters} (\textit{ApJL}), \textit{Astrophysical Journal Supplement} (\textit{ApJS}) and \textit{Astronomical Journal} (\textit{AJ}).} (\textit{AAS journals}), have in fact abandoned the physically printed form altogether. It has also become usual for articles submitted to peer-reviewed publications to be uploaded to the \textit{arXiv} online pre-print server. The emergence of \textit{supplementary online material}, such as extensive data tables, is possibly the prime example of the benefits that have resulted (so far) from the implementation of electronic journals. New initiatives such as that of \textit{Research Objects}\footnote{ \url{http://www.researchobject.org}} aim at further exploiting online technologies to group more strongly scientific articles with their associated datasets, tools, workflows and methodologies \citep[e.g.][]{Hettne12,Hettne13} in order to improve their identification, legacy and re-usability \citep[see also][]{Goodman15}.

Given a largely electronic publishing landscape, one may wonder why no well-established alternative(s) to the classical 2-D printable diagrams exist at this time. The problem of sharing and publishing multi-dimensional datasets has of course been identified in the past. Besides movies and animations, several alternative solutions have been discussed in recent years in the field of Astrophysics, including interactive {\pdf} documents \citep{Barnes08}, stereo pairs \citep{Vogt12}, augmented reality \citep{Vogt13}, 3-D printing \citep{Steffen14,Vogt14,Madura15} and interactive {\html} documents \citep{Vogt14}. However, most of these propositions suffer from two distinct and equally severe flaws that hinder their expansion in the field of Astrophysics. 

First, these techniques have been largely associated with specific software, some of which not open-source. In the case of the interactive {\pdf} approach, \cite{Barnes08} suggested the \textsc{s2plot} software \citep{Barnes06} for generating 3-D models, and relied on the commercial \textsc{adobe acrobat professional} software for the conversation of the model to the {\ud} file format: the only 3-D file format suitable for the inclusion of interactive 3-D models in {\pdf} documents. Relying on specific software restricts the \textit{freedom of choice} of the user, and often impedes on one's ability to experiment with these new techniques. Software is also subject to very rapid changes and evolution, so that specific tools can become obsolete unexpectedly rapidly. In the case of interactive {\pdf} for example, \textsc{adobe acrobat professional} stopped supporting the inclusion of {\ud} files in interactive documents in v10; a task that was then relegated to third party software.

Second, the degree to which these different methods are supported by the principal scientific journals in the field of Astrophysics varies widely, and no technique is currently ``actively'' supported and encouraged by all journals. The different visualization techniques mentioned above are therefore often regarded as \textit{experimental}. The need for specific software to implement these different solutions certainly also reinforces this impression.

The present article introduces the concept of the ``{\xd} pathway'' as a new approach to visualizing, sharing and publishing multi-dimensional datasets that solves the above mentioned issues. Specifically, the {\xd} pathway suggests the utilization of the {\xd} file format as a centralized mean of implementing advanced visualization solutions via 3-D diagrams for astrophysical multi-dimensional datasets.

The primary goals of this article are: 1) to further demystify the use of advanced visualization techniques in the field of Astrophysics, 2) to introduce the {\xd} pathway as a viable and polyvalent approach to advanced multi-dimensional data visualization and 3) to provide a full set of dedicated astrophysical examples (including complete scripts, datasets and step-by-step instructions) to enable the interested readers to familiarise themselves and implement the {\xd} pathway rapidly and easily with their own datasets.

This article is structured as follows. The notion of the {\xd} pathway is introduced in Section~\ref{sec:x3d_pathway} and example implementations are presented in Section~\ref{sec:implementing}. The demonstration datasets are described in Section~\ref{sec:data}. Each branch of the {\xd} pathway is then discussed individually, including: the creation of {\xd} models in Section~\ref{sec:python}, the creation of interactive {\html} documents in Section~\ref{sec:html}, the connection of {\xd} to 3-D printing in Section~\ref{sec:3dprinting} and the creation of high quality movies and animations in Section~\ref{sec:movies}. We present our conclusions in Section~\ref{sec:conclusions}. All uniform resource locators (URLs) provided in this article are valid as of the date of submission. Wherever available, digital object identifiers (DOI) are quoted instead.
\\

\textit{Note to arXiv readers: although some DOIs remain to be minted by ApJ upon publication of the article, all the different scripts, instructions sets, movies, etc ... mentioned in the text are already available in a dedicated Github repository. See \url{https://github.com/fpavogt/x3d-pathway}, and/or DOI:\href{http://dx.doi.org/10.5281/zenodo.31953}{10.5281/zenodo.31953}. For yet un-minted DOIs, a temporary URLs is provided instead.}
 
 \section{The {\xd} pathway}\label{sec:x3d_pathway}

The {\xd} (Extensible 3-D) file format is an International Standards Organization (ISO) ratified standard developed and maintained by the Web3D Consortium\footnote{\url{http://www.web3d.org}}. A detailed description of this file format is provided by \cite{Daly07}, to which we refer the interested reader for further information. Repeating verbatim the description provided by the Web3D Consortium, ``\textit{{\xd} is a royalty-free open standards file format and run-time architecture to represent and communicate 3-D scenes and objects using \textsf{XML}}''. The {\xd} file format has evolved from the Virtual Reality Modeling Language (\textsf{VRML}) format, with the intent to become the standard file format for the publication of 3-D graphics on the World Wide Web. While the {\xd} standard is still evolving, backwards compatibility of the file format is a key objective pursued by the Web3D Consortium \citep{Daly07}. The Web3D Consortium has also established connection with the World Wide Web Consortium\footnote{ W3C; \url{http://www.w3.org/}} and other partners\footnote{ \url{http://www.web3d.org/about/liaisons}} to ensure the compatibility of the {\xd} file format with the evolving World Wide Web.

The {\xd} file format uses a rectangular, right-handed coordinate system. It supports (among other elements) polygonal and parametric geometry, with data points and meshes represented via solid shapes and surfaces. The {\xd} standard also supports the inclusion of 2-D text in 3-D space, useful for annotating models, as well as lighting elements and materials. Volumetric rendering (with an emphasis on human anatomy representation) is being developed by the \textit{Medical Working Group for {\xd}}\footnote{ \url{http://www.web3d.org/working-groups/medical}}, with dedicated (and still limited) specifications included in ISO {\xd} v3.3. Several quality assurance tools developed alongside the {\xd} standard itself, for example {\xd}\textsc{-Edit}\footnote{ \url{https://savage.nps.edu/X3D-Edit}} or the online {\xd} \textsc{Validator}\footnote{ \url{https://savage.nps.edu/X3dValidator}} allow to easily inspect {\xd} files to ensure they contain no syntax errors and follow the {\xd} file format conventions.

\begin{figure*}[htb!]
\begin{center}
\tikzstyle{every node}=[font=\normalsize]
\tikzstyle{gcircle}=[rectangle, very thick, minimum size=0.5cm,draw=olive!100,fill=green!20,text centered,rounded corners, text centered]
\tikzstyle{bsquare}=[rectangle,very thick,minimum size=0.5cm,draw=black!100,fill=black!5,text centered,]
\tikzstyle{bdsquare}=[rectangle,dotted,very thick,minimum size=0.5cm,draw=black!100,fill=black!5,text centered,]
\tikzstyle{osquare}=[rectangle,very thick,minimum size=0.5cm,draw=orange!100,fill=black!5,text centered,]
\tikzstyle{pellipse}=[rectangle,very thick,minimum size=0.5cm,draw=purple!100,fill=purple!5, rounded corners,text centered]
\begin{tikzpicture}[auto, outer sep=3pt, node distance=3.5cm and 2cm,>=triangle 45]
\node [gcircle, align=center] (start) {\textbf{START }};
\node [bsquare, below =0.5cm of start] (fits) {\textbf {\fits\ $|$ \txt\ $|$ $\ldots$\,}};
\node [osquare, right= 3.5cm of fits, align=center] (x3d) {\textbf \xd\,};
\node [bsquare, right= 4.5cm of x3d] (stl) {\textbf \stl\,};
\node [bsquare, above= 2cm of stl] (png) {\textbf {{\png}s $|$ \textsf{MP4} $| \ldots$}\,};
\node [bsquare, below=2cm of stl] (html) {\textbf \html\,};
\node [bsquare, below=2cm of  html] (pdf) {\textbf \pdf\,};
\node [bsquare, left= 4.4cm of pdf] (u3d) {\textbf \ud\,};
\node [bdsquare, above left=1.5cm and 1.5cm of  u3d] (wrl) {\textbf \wrl\,};
\node [pellipse, right of=  png] (movie) { \textbf{$^\text{movies \&}_\text{animations}$}};
\node [pellipse, right of= stl] (printing) { \textbf{$^\text{\ \ 3-D}_\text{printing}$}};
\node [pellipse, right of=  html] (doc) { \textbf{$^\text{interactive}_\text{documents}$}};
\node [pellipse, right of=  pdf] (doc2) { \textbf{$^\text{interactive}_\text{documents}$}};
\draw [->,thick,double equal sign distance, >=latex',olive] (start) to node {} (fits) ;
\draw [->, very thick,orange] (fits) to node [black]{\scriptsize \python\ (\mayavi)} (x3d) ;
\draw [->, very thick,orange] (x3d) to node[black,midway]{\scriptsize\blender, \meshlab$^\text{\ \danger}$} (stl) ;
\draw [->, very thick,orange, bend left] (x3d.45) to node[anchor=center,midway,black, fill=white]{\scriptsize \blender} (png.west) ;
\draw [->, very thick,orange, bend right] (x3d.315) to node[pos=0.35,anchor=center,black,fill=white]{\scriptsize Text editor}(html.west) ;
\draw [->, very thick,orange,dashed] (x3d) to node [anchor=center,pos=0.55,black, fill=white,text width=3cm,align=center]{\scriptsize \pdfrg$^{\ \$}$ \meshlab$^{\text{\ \danger}}$} (u3d) ;
\draw [->, very thick,orange,dashed] (u3d) to node [black,align=center,midway]{\scriptsize \LaTeX\ (\textsc{media9}; \textsc{\st{movie15}})} (pdf) ;
\draw [->,thick,dotted,bend left] (fits.south east) to node[black,fill=white,anchor=center,midway] {\scriptsize \python\ (\mayavi)} (wrl.north) ;
\draw [->,thick,dotted,bend right] (wrl.south) to node[black,fill=white,anchor=center,pos=0.3,align=center,text width=3cm] {\scriptsize \pdfrg$^{\ \$}$ \meshlab$^{\text{\ \danger}}$} (u3d.west) ;
\draw [->,thick, bend left, dotted] (fits.north east) to node[black,fill=white,anchor=center,pos=0.37,text width=2.5cm,align=center] {\scriptsize \centerline \python\ \newline(\textsc{matplotlib}, \mayavi, \yt, \astroblend)} (png.170);
\draw [<->,thick,double equal sign distance, >=latex',purple] (stl) to (printing);
\draw [<->,thick,double equal sign distance, >=latex',purple] (png) to (movie);
\draw [<->,thick,double equal sign distance, >=latex',purple] (html) to (doc);
\draw [<->,thick,double equal sign distance, >=latex',purple] (pdf) to (doc2);
\end{tikzpicture}
\vspace{15pt}
\caption{Schematic of the {\xd} pathway, illustrating the routes leading to the creation of different advanced visualization techniques. The {\xd} file format forms the core of a product tree (thick orange arrows) that includes (but is not limited to) interactive {\html} documents, 3-D printing and high-end animations \& movies. The {\xd} file format can also lead to interactive {\pdf} documents via a conversion to the specific and dedicated {\ud} file format which uniquely serves that purpose. This ``{\xd} to {\pdf}'' branch is marked with dashed arrows, as it (currently) requires either commercial software (marked with $^{\$}$) or open source software with significant limitations (marked with $^\text{\danger}$). The {\xd} pathway is based on file formats, and software links in the diagram only list our (current) personal favoured solutions discussed in this article. Dotted arrows illustrate the alternative route to the creation of interactive {\pdf} models implemented by \cite{Vogt15}, as well as a more direct (and usual) route to the creation of movies and animations by-passing the {\xd} file format. }\label{fig:x3d_pathway}
\end{center}
\end{figure*}
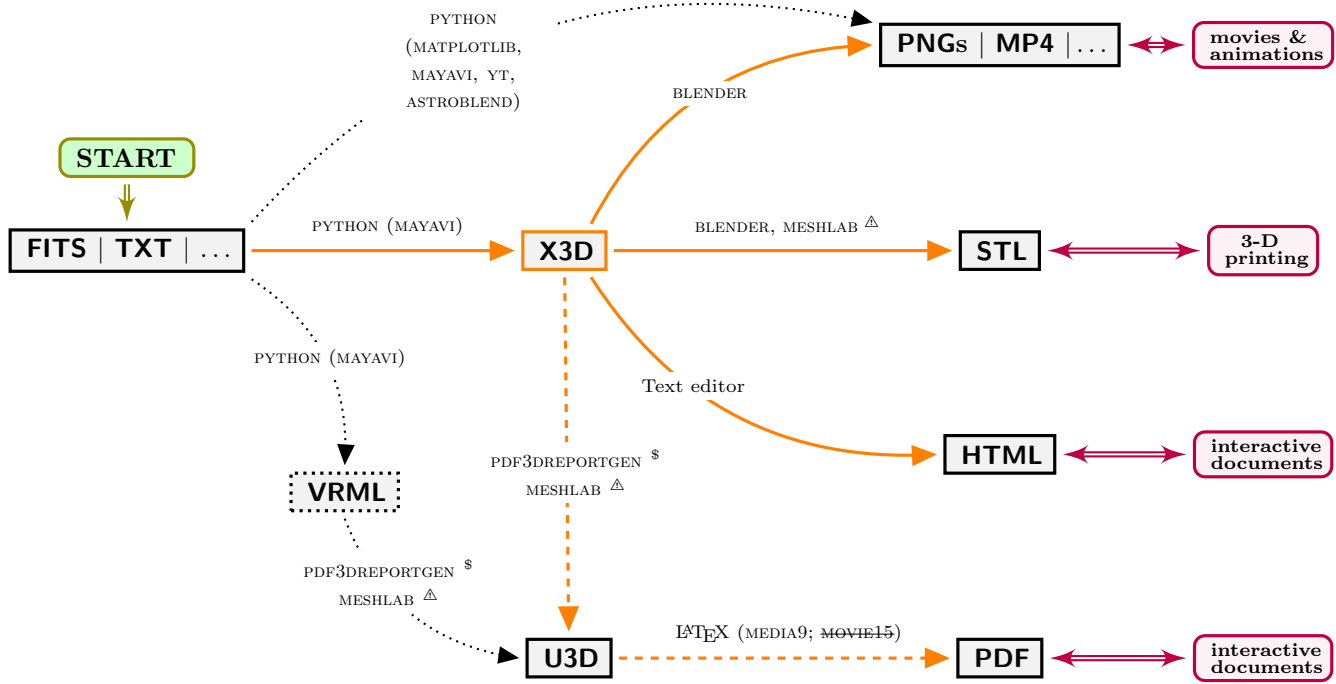

The underlying idea of the {\xd} pathway is that the {\xd} file format lies at the core of different visualization solutions for multi-dimensional datasets via 3-D diagrams, including interactive {\html} documents, 3-D printing and high-end movies and animations. A schematic guide to the implementation of these advanced visualization techniques (via the {\xd} standard) for multi-dimensional datasets is presented in Figure~\ref{fig:x3d_pathway}. The links forming the {\xd} pathway are marked with thick orange connections.

The {\xd} pathway is not software dependant. The software names connecting each nodes of the pathway in Figure~\ref{fig:x3d_pathway} are existing and working solutions which we believe are currently the most suitable for astronomers with no prior experience in 3-D data visualization, and for which we provide dedicated examples as supplementary material to this article. But a wide range of software compatible with the {\xd} pathway exits, some intrinsically more complex and/or non-open source, but possibly more capable and/or familiar to specific users. Specific alternatives will be mentioned in Sections~\ref{sec:python} to \ref{sec:movies}.

Rather than specific software, the {\xd} pathway is centred on file formats - and in particular on the {\xd} file format. Although file formats are not immune to change and evolution, such changes can be expected to be gradual and slow in comparison to the evolution of associated software. In fact, relying on file formats rather than specific software not only ensures the long-term stability of the {\xd} pathway, but also does not restrict the existing (or future) \textit{freedom of choice} of astronomers to implement it. The {\xd} standard also brings the important features of \textit{re-usability} and \textit{lasting legacy} of the material, as backward compatibility is a key element of the {\xd} file format set by the Web3D consortium. Finally, the {\xd} standard is fully open source, a fact which acts as a guarantee that we (as a community) are able to directly influence its evolution to match our needs; for example via the hypothetic creation of a new Web3D working group\footnote{ \url{http://www.web3d.org/working-groups}} dedicated to the promotion and development of new {\xd} specifications for the field of Astronomy and Astrophysics.

\subsection{Interactive {\pdf} and interactive {\html} documents}

Interactive {\pdf} documents are a \textit{reader-friendly} solution to sharing interactive 3-D models with the scientific community, as they do not require readers to have access to costly commercial software\footnote{ Interactive {\pdf} documents can be opened using recent versions of \textsc{adobe acrobat reader}, freely available online: \url{https://get.adobe.com/reader/}}. Interactive {\pdf} articles have been used successfully be several authors in recent years in the field of Astrophysics \citep[see e.g.][]{Springbob12, Putman12, Springob14, Miki14, Steffen14, Madura15, Vogt15}. Of course, one of the main advantage of interactive {\pdf} documents (from a scientist perspective) is that interactive models can be embedded directly inside the {\pdf} of an article, thereby ensuring their long term accessibility. One pathway  \citep[as implemented by][]{Vogt15} leading to the {\ud} file required for the creation of an interactive {\pdf} document is illustrated in Figure~\ref{fig:x3d_pathway}.  

But interactive {\pdf} documents also have two distinct drawbacks. First, interactive {\pdf} documents can (at the time of submission of this article) only be opened using \textsc{adobe acrobat reader} v9.0 or above, as other {\pdf} reading software (such as Apple's \textsc{preview}) will only display a still cover image. As such, interactive {\pdf} documents very strongly restrict the freedom of choice of the reader. Second, the {\ud} file format is non-straightforward to generate. For example, \cite{Vogt15} used the commercial software {\pdfrg}. {\meshlab} \footnote{ \url{http://meshlab.sourceforge.net/}} is an open-source alternative often cited as the solution to the creation of {\ud} files. In reality, {\meshlab} \textsc{(v1.3.3)} is suitable for creating {\ud} files only under very specific circumstances, as this software currently does not support semi-transparent layers, and is unable to conserve a complex model tree throughout the conversion process.

The idea of interactive {\html} documents was recently used by \cite{Vogt14} to publish an interactive model of a 3-D emission line ratio diagram. Unlike interactive {\pdf} documents, interactive {\html} documents do not strongly restrict the user's freedom of choice (they are compatible with all major web browsers, see Section~\ref{sec:html}), and most importantly, the {\xd} file format (at the base of interactive {\html} documents) can be freely created using open-source software (e.g. {\python}). These publishing advantages are clear enough that the \textit{AAS journals} are now actively supporting and promoting the use of interactive {\html} documents to share multi-dimensional datasets, where Figure~1 in \cite{Vogt14} was used as a proof-of-concept\footnote{ \href{https://dx.doi.org/doi:10.1088/0004-637X/793/2/127/data}{DOI:10.1088/0004-637X/793/2/127/data}}. Interactive 3-D graphics are but one of a suite of plans for enhanced graphics and advances in publishing technology currently being discussed by the \textit{AAS journals}\footnote{ http://aas.org/posts/news/2015/02/changes-ahead-aas-journals}.

Although we believe that the intrinsic advantages of interactive {\html} documents render them more suitable for the field of Astrophysics than interactive {\pdf} documents, we stress that these solutions are not mutually exclusive. Indeed, {\ud} files can be generated from {\xd} files (see Figure~\ref{fig:x3d_pathway}), so that interactive {\pdf} documents in fact represent another branch of the {\xd} pathway tree. We shall not discuss interactive {\pdf} in details in this article, as 1) their creation (using the \textsc{media9} and/or the now obsolete \textsc{movie15} \LaTeX packages) and features have already been discussed by \cite{Barnes08}, and 2) to the best of our knowledge the creation of {\ud} files remains a difficult problem to which no real open-source solution currently exists (modulo {\meshlab} under specific circumstances).

\section{Implementing the {\xd} pathway}\label{sec:implementing}

In this Section, we implement a practical demonstration of how the {\xd} pathway can be used to aid in the visualization and publication of complex, multi-dimensional datasets in Astrophysics. Although the {\xd} pathway does not rely on specific software \textit{per se}, software is certainly still required to exploit and implement it. Hence, this Section is also intended as a mean of providing complete and realistic scripts and step-by-step instructions to the interested reader. All the scripts are available in a dedicated Github repository\footnote{ \url{https://github.com/fpavogt/x3d-pathway}}, and for legacy purposes, a DOI was tied to the release v0.9 of the repository \citep[DOI:\href{http://dx.doi.org/10.5281/zenodo.31953}{10.5281/zenodo.31953};][]{Zenodo_0.9} using Zenodo. We set our focus on two specific tools: the {\mayavi}\footnote{ Throughout this article, when we refer to {\mayavi}, we mean {\mayavi}2, the second generation module developed and maintained by Enthought: \url{http://docs.enthought.com/mayavi/mayavi/}} module in {\python} \citep[][used for the initial 3-D plotting of multi-dimensional datasets and the export to the {\xd} file format]{Ram11}, and {\blender}\footnote{ \url{http://www.blender.org}} (for the creation of high-end movies \& animations and 3-D printable models).

The structure and content of the material inside the v0.9 release of the Github repository associated to this article is described in detail in Appendix~\ref{app:supp}. The different scripts and sets of instructions provided in the repository are not discussed in details in the core of this article, aside of a general description, as such a description is not of direct importance to the concept of the {\xd} pathway itself. Furthermore, these scripts represent only one (among a few equivalent) ways to implement the {\xd} pathway, so that describing specific examples in details would unjustifiably emphasize specific software. 

\subsection{Demonstration datasets}\label{sec:data}

\subsubsection{Green and red dices}

\begin{figure}[htb!]
\centerline{\includegraphics[scale=0.4]{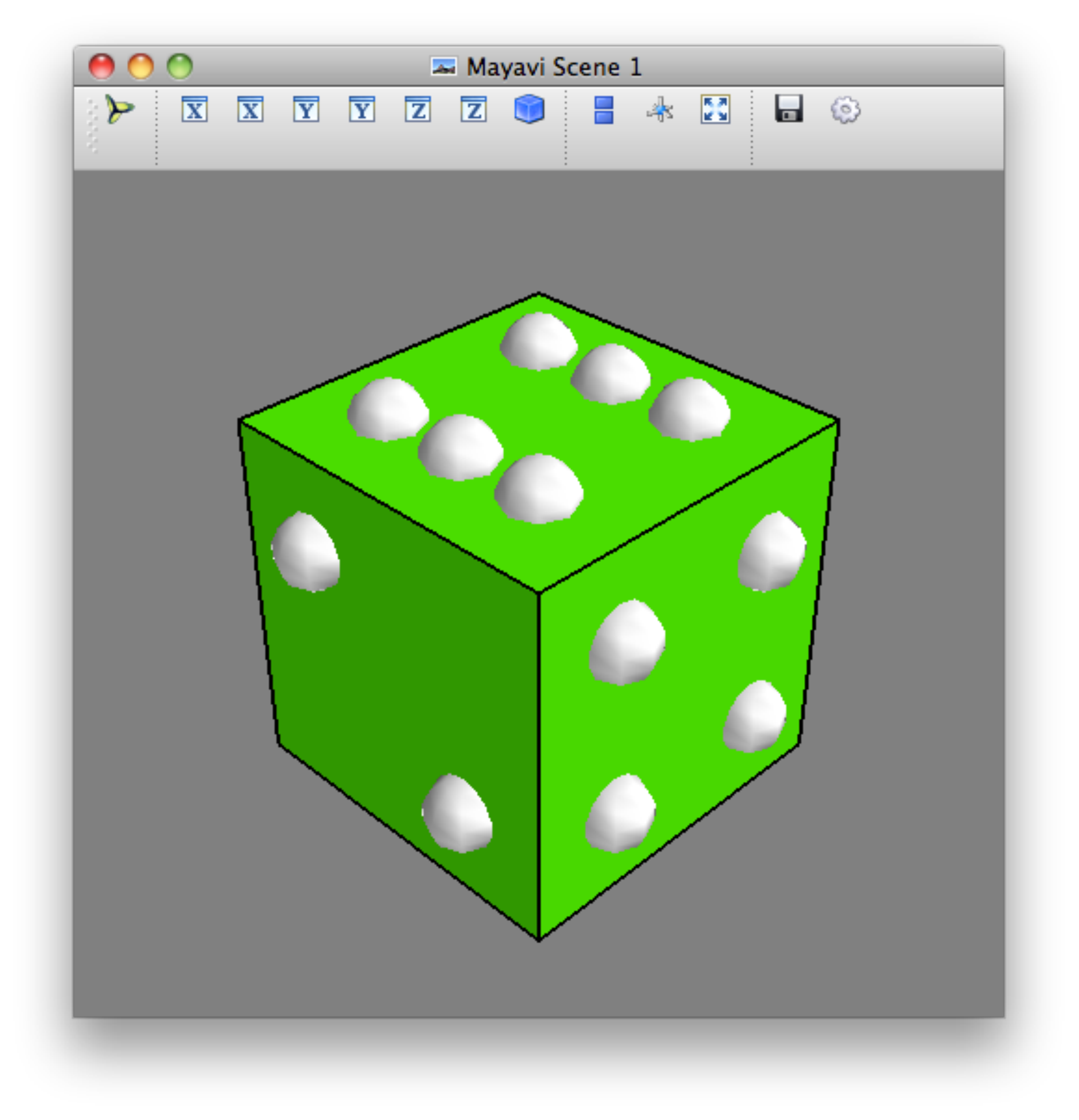}}
\centerline{\includegraphics[scale=0.4]{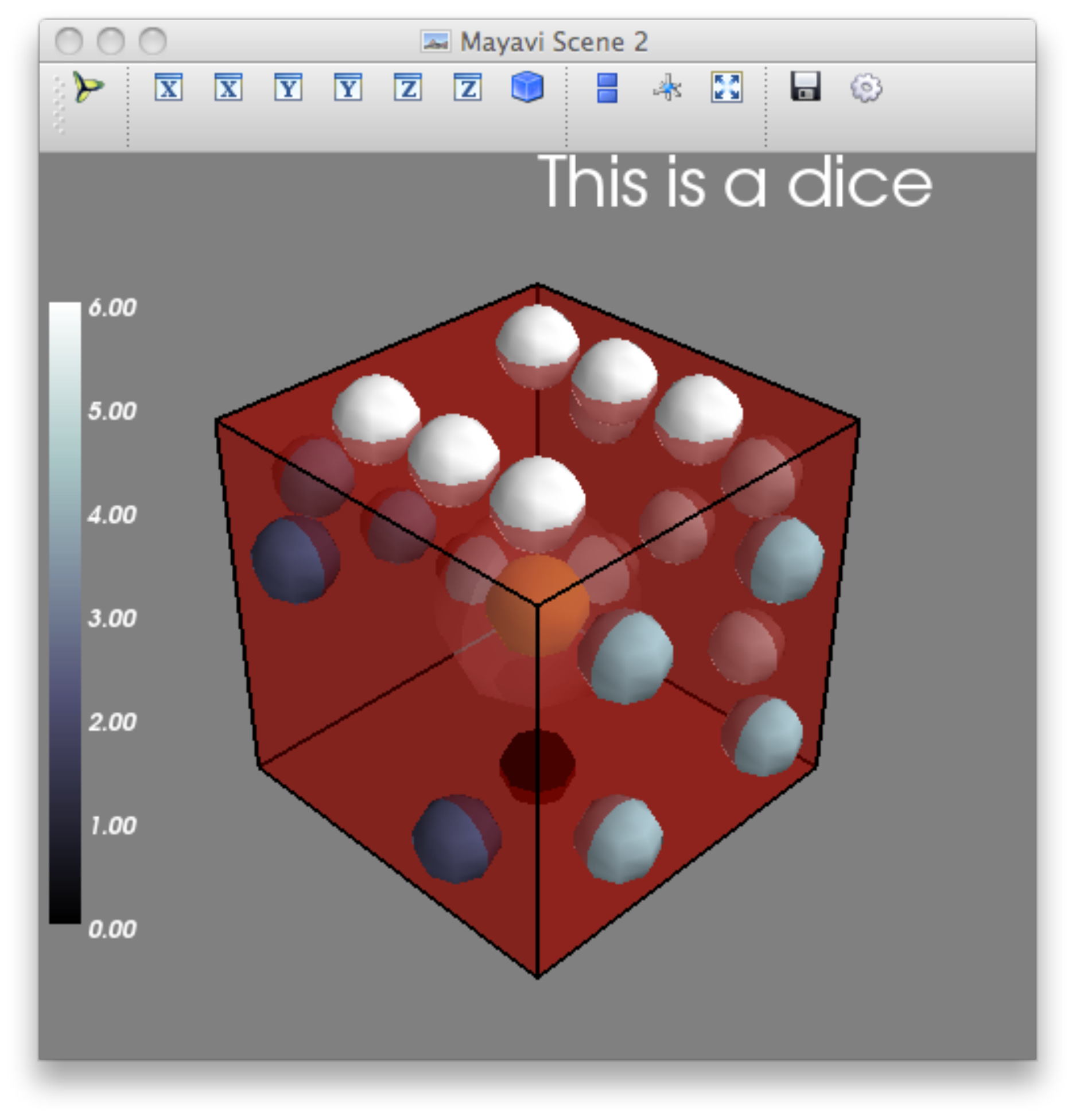}}
\caption{The green (top) and red (bottom) dices used to demonstrate how to implement the {\xd} pathway. These models complement the more realistic example of the \textit{VLA} observations of the {\HI} gas inside HCG~91 by providing two additional examples of incremental complexity. The interactive version of these diagrams is accessible via the DOI: \href{https://dx.doi.org/TBD}{TBD}. \textit{For arXiv readers: the interactive diagrams are available here until publication: \href{http://www.mso.anu.edu.au/\~fvogt/website/arXiv/x3d/red\_dice.html}{http://www.mso.anu.edu.au/$\sim$fvogt/website/arXiv/ x3d/red\_dice.html}}. }\label{fig:dices}
\end{figure}

In order to provide series of examples of increasing complexity, we first use two basic 3-D structures: a green dice and a red dice. These models are designed with incremental complexity leading to the visualization of a datacube from the \textit{Very Large Array} (\textit{VLA}) (see Section~\ref{sec:vla}). Screenshots of both the green and red dice examples (as drawn inside the {\mayavi} interactive plotting window) are presented in Figure~\ref{fig:dices}.

The green dice is our most simple example. It illustrates the basic, minimal steps required to visualize multi-dimensional data interactively using {\mayavi}, publish it using interactive {\html} documents, and animate it using {\blender}. The red dice example contains a semi-transparent cube with two inner-spheres (white and yellow), some 3-D text and a colorbar. This example was designed to hit some of the current limitations associated with creating {\xd} files using {\mayavi} while keeping the actual 3-D structure as simple as possible. 

\subsubsection{The {\HI} gas content of HCG~91}\label{sec:vla}

Our realistic astrophysical example is based on the \textit{VLA} observations of the {\HI} gas distribution and kinematics around the galaxies inside the Hickson Compact Group (HCG) 91. The complexity of this \textit{VLA} dataset makes it a prime example to illustrate how advanced visualization techniques such as interactive {\html} documents, 3-D printing, and high-end animations (implemented via the {\xd} pathway) can help in visualizing, understanding, sharing and publishing the structural content of the data. Visualizing the cold gas distribution in and around galaxies to trace their evolution is also one of the main scientific drivers of the \textit{SKA}, and 3-D diagrams can play an important role in the analysis of these datasets \citep[see e.g.][]{Punzo15}. 

HCG 91 is a group of 4 galaxies first catalogued by \cite{Hickson82} at a distance of $\sim100$\,Mpc in the constellation of Piscis Austrinus. The {\HI} cold gas distribution and kinematics in this group, which is a prime tracer of ongoing gravitational interactions between its member galaxies and of the overall evolutionary stage of the group \citep[e.g.][]{Verdes-Montenegro01} was observed by the \textit{VLA} in New Mexico (P.I.: L. Verdes-Montenegro, P.Id.: AV0285) on 2005 October 5. This dataset was used by \cite{Vogt15} to characterize the evolutionary stage of the galaxy HCG~91c, and will be analyzed in details in Verdes-Montenegro et al. (in prep.) to which we refer the reader for further details on the scientific implications of these observations. Here, we restrict ourselves to a generic description of the data, which is used as a realistic and representative example of multi-dimensional astrophysical datasets.

\begin{figure*}[p!]
\centerline{\includegraphics[scale=0.37]{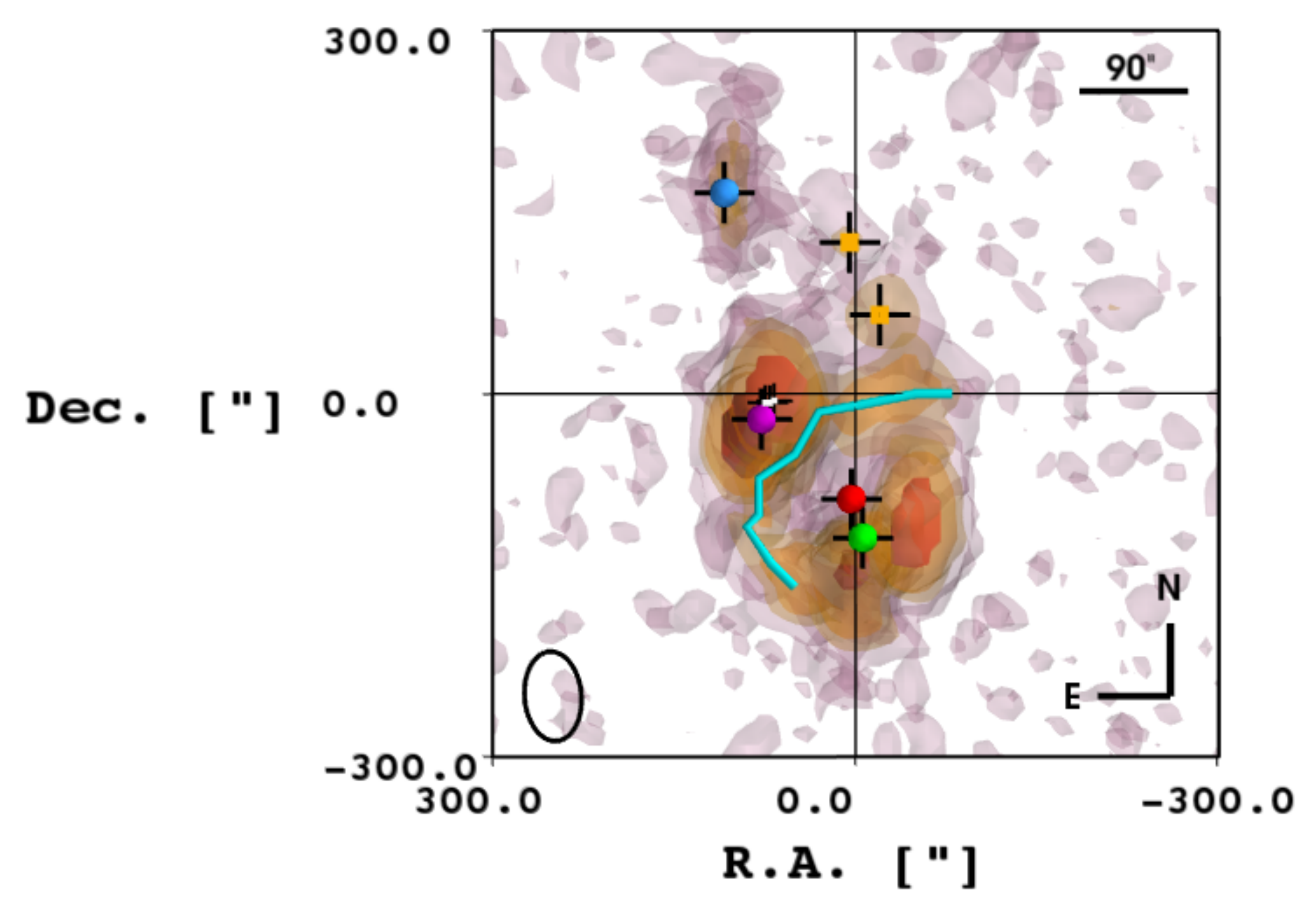}}
\centerline{\includegraphics[scale=0.37]{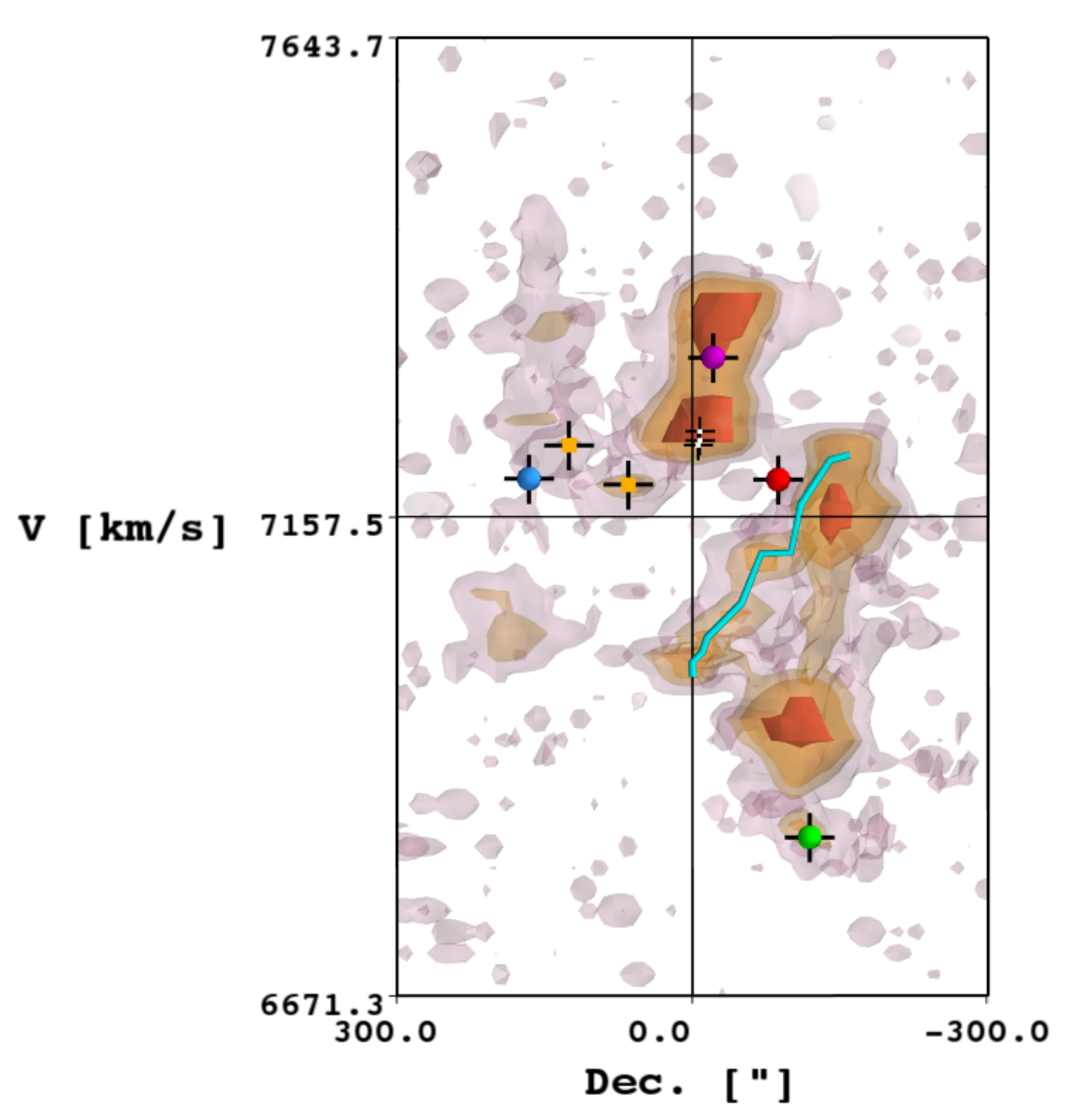}\quad\includegraphics[scale=0.37]{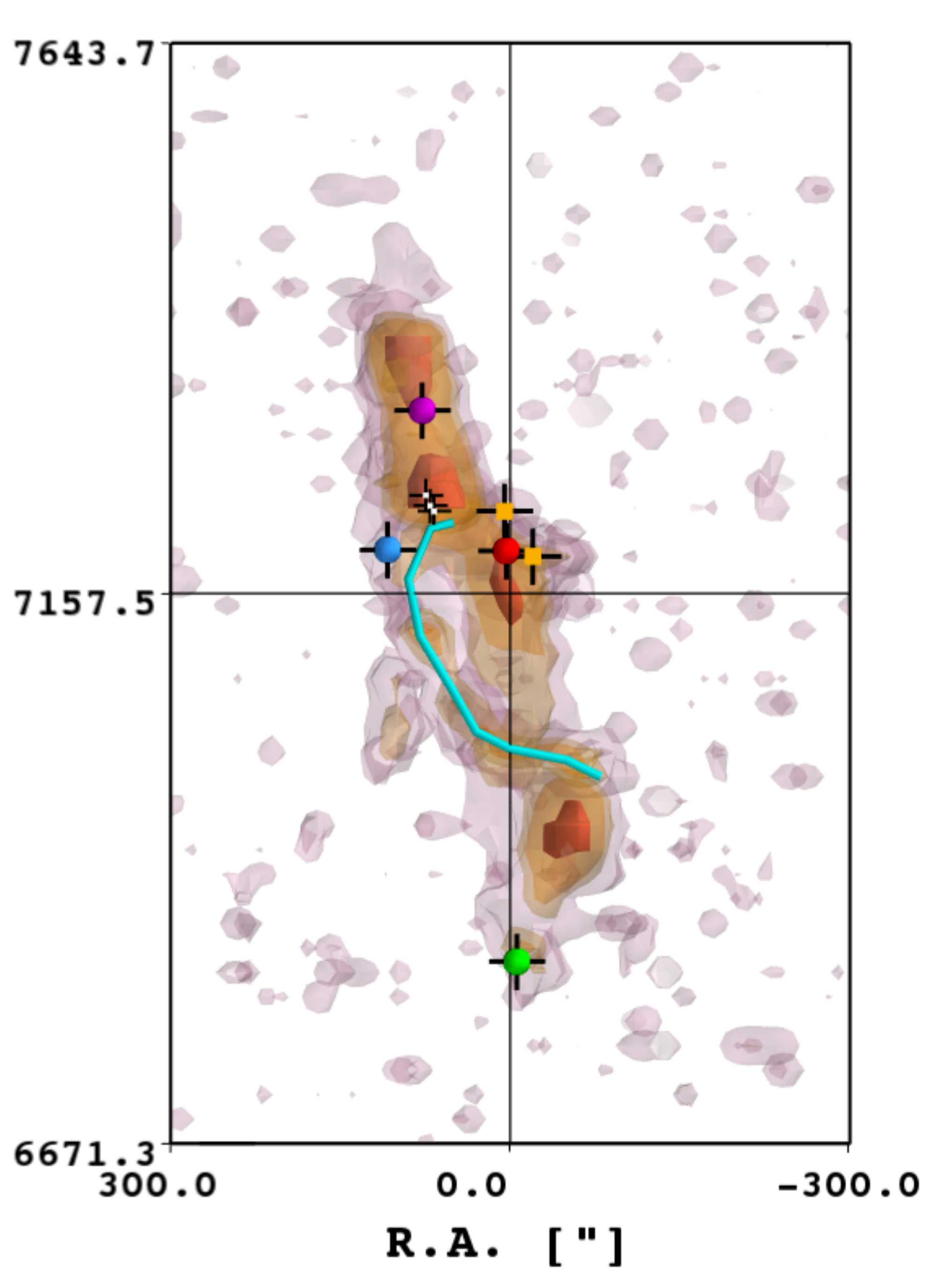}}

\caption{The {\HI} gas distribution and kinematics of the compact group of galaxies HCG~91 observed by the \textit{VLA}. Individual intensity iso-contours at 1.3, 2.5, 3.5 and 6.0 mJy/beam are fitted in 3-D inside the datacube before being projected along top-front-side projections. The inner-most contour is fully opaque, while the others are set at 80\% transparency to enable a \textit{see-through} effect. The galaxies HCG~91a, b, c and d are marked with green, blue, purple and red spheres, respectively. A large tidal tail originating to the South-East of HCG~91a is traced inside the 3-D cube with a cyan line, and other elements of interest are marked with white and yellow cubes \citep[see][for details]{Vogt15}. In the top panel, the black ellipse traces the beam size and orientation associated with the \textit{VLA} data. The interactive version of this Figure is accessible via the DOI:\href{https://dx.doi.org/TBD}{TBD}.  \textit{For arXiv readers: the interactive diagram is available here until publication: \href{http://www.mso.anu.edu.au/\~fvogt/website/arXiv/x3d/HCG91.html}{http://www.mso.anu.edu.au/$\sim$fvogt/website/arXiv/x3d/HCG91.html}}. }\label{fig:mayavi}
\end{figure*}

The datacube is in units of ([X]: arcsec; [Y]: arcsec; [v]: km\,s$^{-1}$). A subset of the complete \textit{VLA} cube, spanning 10\,arcmin\,$\times10$\,arcmin\,$\times\,972.4$\,km\,s$^{-1}$ and centered on HCG~91, is shown as a top-front-side projection triplet in Figure~\ref{fig:mayavi}. Three semi-transparent and one opaque iso-contours are used to reveal the complex structure of the {\HI} gas emission in this spatial-kinematic volume, alongside colored markers indicating the location of features of interest. In particular, the cyan line traces a tidal tail wrapping around the main group galaxy HCG~91a (marked by a green sphere) with a velocity extent of $\sim400$\,km\,s$^{-1}$.

\subsection{Creating {\xd} files}\label{sec:python}

There exists many different software solutions to create {\xd} files given the central role that this data format is gaining on the World Wide Web scene for sharing 3-D graphics\footnote{See e.g. \url{http://www.web3d.org/x3d/content/examples/X3dResources.html\#Applications}}.  Here, we rely on the {\mayavi} module in {\python} \citep{Ram11} for three reasons. 

First, the {\python} programming language has been steadily growing in popularity within the Astrophysical community in recent years. A {\python} solution to the problem of multi-dimensional data visualization will therefore be best suited to many astronomers. Second, {\mayavi} offers a simple and rapid way to generate {\xd} files for non-experts with no \textit{a priori} knowledge of the package. In particular, the {\mayavi} syntax is extremely reminiscent of the popular {\matplotlib} package \citep{Hunter07} dedicated to the creation of 2-D diagrams in {\python}. Third, {\mayavi} intrinsically provides an interactive view of the dataset without the need to export it first. In other words, {\mayavi} is perfect to allow individual researchers to rapidly explore the content of a multi-dimensional dataset interactively and assess the need for and eventual benefit of additional forms of 3-D data visualization (e.g. in the form of interactive {\html} documents) without having to implement them first. Detailed examples of the {\mayavi} capabilities and associated syntax are accessible online\footnote{ \url{http://docs.enthought.com/mayavi/mayavi/auto/examples.html}}, to which we refer the interested reader for more information. 

We do provide in the Github repository associated to this article the full {\python} scripts used to generate the 3-D models of the green dice, red dice and {\HI} datacube from the \textit{VLA} (see Appendix~\ref{app:supp} and DOI:\href{http://dx.doi.org/10.5281/zenodo.31953}{10.5281/zenodo.31953}). Figure~\ref{fig:mayavi} was generated using a slightly modified version of the latter script. We note that our examples do not explore the issue of mesh reduction or decimation, which may be required for specific models/structures before generating {\xd} files.

The ease of use of {\mayavi} (and even more so for existing {\python} users) is counter-balanced by specific drawbacks (e.g. the currently-imperfect underlying \textsc{Visualization Toolkit (VTK) {\xd} exporter v0.9.1}). Both the red dice and HCG~91 examples are specifically designed to hit the existing limitations of {\mayavi} \citep[see also][]{Punzo15}, and offer possible solutions, including the creation of axes with ticks and labels, and/or the addition of a colorbar to the scene. The example scripts are clearly commented regarding these aspects, which we voluntarily do not discuss further in this article. 

\cite{Punzo15} presented a detailed comparison between {\mayavi} and other software (\textsc{paraview}, \textsc{3dslicer} and \textsc{imagevis3D}) with similar \textit{visualization} capabilities \citep[see also][regarding \textsc{frelled}]{Taylor15}. Among these, only \textsc{paraview}\footnote{ \url{http://www.paraview.org}} \citep{Ahrens05} is able to export models to the {\xd} file format. The \textsc{yt project}  \citep{Turk11} represents another open-source alternative to {\mayavi} to create {\xd} models.

\subsection{From {\xd} to interactive {\html} documents}\label{sec:html}

The most straightforward application for {\xd} files (because of their design driver) is their inclusion in {\html} documents on the World Wide Web. This article is not the first time that the use of the {\xd} file format for that specific purpose has been advocated in the field of Astrophysics. \cite{Fluke09} already suggested the use of the {\xd} file format, in combination with a dedicated \textsc{Flash} plugin, for visualizing data products from the \textsc{s2plot} software \citep{Barnes06}. However, the need for a specialized \textsc{Flash} plugin to access their interactive 3-D models on the web, albeit open-source, remained a negative factor affecting the expansion of this technology. 

Fortunately, as part of the {\xd} version 4.0, full {\html}5 support using the {\xdom} approach \citep{Behr09,Behr10} is being implemented inside the {\xd} standard. In other words, interactive {\xd} models ought to be accessible directly (without the need for external plugins) by all web browsers in the near future. At the time of publication of this article, interactive 3-D models placed within {\html} pages using the {\xdom} approach can already be loaded directly (without the need for any external plugins) with up-to-date versions of the {\firefox}, {\chrome}, {\safari} and {\ie} web browsers that are webGL-enabled\footnote{A detailed and up-to-date web browser compatibility list with the {\xdom} approach is available online: \url{http://www.x3dom.org/?page_id=9}}. 

The interactive {\html} counterparts to Figures~\ref{fig:dices} and \ref{fig:mayavi} are accessible using the respective DOIs: \href{https://dx.doi.org/TBD}{TBD} and \href{https://dx.doi.org/TBD}{TBD}. Software-wise, the creation of interactive {\html} documents only requires a text editor to create the {\html} script and a 3-D model as an {\xd} file. \\

\textit{For arXiv readers: the interactive diagrams are available here until publication: \href{http://www.mso.anu.edu.au/\~fvogt/website/arXiv/x3d/red\_dice.html}{http://www.mso.anu.edu.au/$\sim$fvogt/website/arXiv/ x3d/red\_dice.html} and \href{http://www.mso.anu.edu.au/\~fvogt/website/arXiv/x3d/HCG91.html}{http://www.mso.anu.edu.au/$\sim$fvogt/ website/arXiv/x3d/HCG91.html} }.\\

We provide in the Github repository associated to this article a detailed walkthrough leading to the creation of the interactive {\html} counterpart of Figures~\ref{fig:dices} and \ref{fig:mayavi}. The example {\html} scripts also detail the creation of interaction buttons to set pre-defined views or display/hide specific datasets. Interaction buttons provide authors with an efficient tool to focus the reader's attention to specific aspects of the interactive model without restricting the reader's freedom to explore the content of the dataset. In the case of the {\HI} gas location and kinematics in HCG~91 for example, the ability to show and/or hide different iso-contours can provide readers with a less obstructed view of the most intense {\HI} emission around each galaxy. We note that our example scripts are applicable for the inclusion of any {\xd} model inside an {\html} document, irrespective of the software used for their creation.

\subsection{From {\xd} to 3-D printing}\label{sec:3dprinting}
The concept of 3-D printing, formally known as additive manufacturing, has been recently suggested and used as a new way to share multi-dimensional datasets in Astrophysics \citep{Steffen14, Vogt14, Madura15}. The emergence and increasing availability of reliable 3-D printers within universities and research institutes in recent years is facilitating researchers' access to this technology, easing experimentation, and reducing the associated costs (both financially and time-wise). 

Clearly, the concept of 3-D printing in Astrophysics is still in its infancy, and its future remains uncertain at this stage. The potential of 3-D printing at a professional level has been explored and discussed by \cite{Madura15}, who found that the technique can help better understand the structural content of complex numerical simulations. Using the physical model of a 3-D emission line ratio diagram created by \cite{Vogt14}, one of us also observed how 3-D printed diagrams can promote and facilitate ``live'' interactions and discussions around a complex dataset between researchers (e.g. in group meetings). 

At the time of submission of this article, the most evident use for 3-D printing in Astrophysics most certainly lies within outreach activities. 3-D printed models provide a new pathway for researchers to share complex datasets and concepts with non-experts. Giving members of the general public the possibility to physically handle real astrophysical datasets enables them to use their kinaesthetic sense (and not solely their sight) to ``visualize'' astrophysical datasets in an un-precedented way. This diversification of senses used to ``perceive'' a given multi-dimensional dataset can lead to an improved understanding of its structure and of its associated parameters. The efforts undertaken at the Space Telescope Science Institute towards sharing \textit{Hubble} images of the Universe with blind and vision-impaired members of the general public using 3-D printing are also particularly noteworthy \citep{Christian14}. 

Whichever future awaits 3-D printing technology in Astrophysics, the {\xd} pathway already provides astronomers with an efficient way to create 3-D printable models of 3-D diagrams. As a well-supported and well-defined 3-D data format, {\xd} models can be easily exported to specific file formats (e.g. {\stl}) required by 3-D printers and associated software. For example, the conversion of an {\xd} file to an {\stl} file can be achieved easily using the open-source software {\meshlab}. Unlike for the creation of {\ud} files, transparency information and complex model trees are not required to be conserved inside {\stl} files, so that {\meshlab} offers a suitable solution in this case. For complex datasets, {\blender} offers a significantly better way to convert 3-D diagrams from {\xd} to 3-D printable models while allowing the addition of support structures to hold free-floating elements. 

\begin{figure*}[htb!]
\centerline{\includegraphics[scale=0.28]{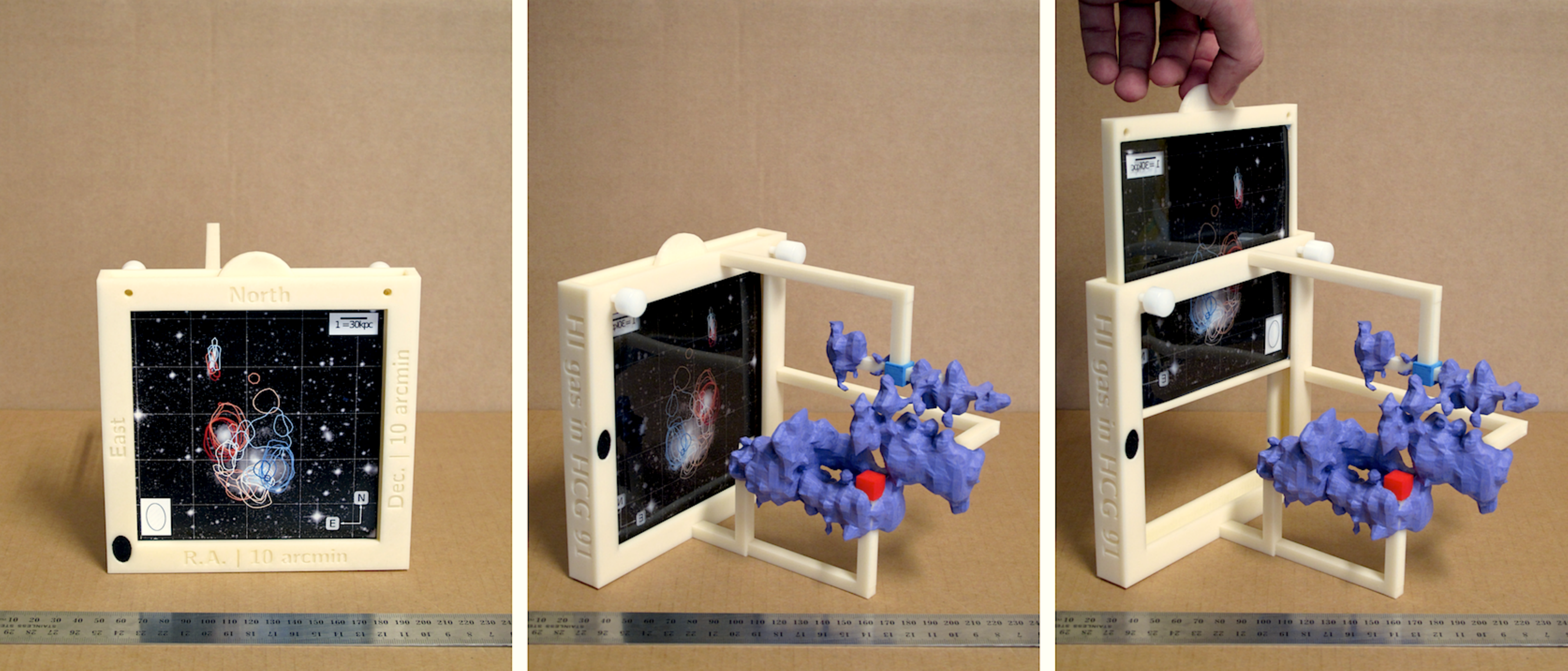}}
\caption{3-D printed model of the {\HI} distribution and kinematics inside HCG~91 (at the 1.3 mJy/beam level). Cubes mark the position of the galaxies in the [X-Y-V] volume, and are colored according to Figure~\ref{fig:mayavi}. The base plate contains a ``sliding-rack'' designed to hold an on-sky picture of the galaxy group, and augment the experience of the 3-D print in allowing a direct comparison between the 2-D projected diagram and the full 3-D structure. Additional information (incl. the size and orientation of the \textit{VLA} beam, visible as a black ellipse in the images) is also engraved in the base plate directly as part of the 3-D printing process. The colors were added manually using acrylic paint, after dissolution of temporary support structures. For reference, the ruler at the bottom of the frames is in cm, and the model has a physical size of [15\,cm\,$\times$\,15\,cm\,$\times$\,19\,cm].}\label{fig:3d_print}
\end{figure*}

We have created a 3-D printable version of the {\HI} gas distribution and kinematics in the compact group of galaxies HCG~91. Two demonstration prints (with a final model size of [15\,cm\,$\times$\,15\,cm\,$\times$\,19\,cm] and [7.5\,cm\,$\times$\,7.5\,cm\,$\times$\,9.5\,cm], respectively) were performed on the \textsf{Fortus 400mc} printer at the Research School of Astronomy and Astrophysics at the Australian National University. The larger print, after dissolution of the temporary support structures and manual reproduction of the color scheme of the model, is presented in Figure~\ref{fig:3d_print}. The corresponding {\stl} file (compatible with most 3-D printer currently available on the market) and the {\blender} file used for the {\xd} to {\stl} conversion process are both included in the Github repository associated to this article.

\subsection{From {\xd} to high-end animations}\label{sec:movies}

Movies and animations have long been used in Astrophysics to share the content of multi-dimensional datasets, where time is being added as an extra-dimension to a given 2-D diagram. Generally speaking, movies consist in a series of still frames (e.g. {\png}s) that are stacked together into an animation, for example using the open-source and polyvalent {\ffmpeg} software. Certainly, the creation of {\png}s does not require the prior creation of an {\xd} model, and can be achieved either using 2-D or 3-D plotting solutions with direct export to still {\png}s, as illustrated in Figure~\ref{fig:x3d_pathway}. 

In this Section, we specifically explore the possibility of implementing high-end animations of 3-D diagrams exported as {\xd} models using {\blender}. {\blender} is a free, open source, multi-platform (Windows, Mac, GNU/Linux) 3-D graphics and animation software package\footnote{\url{http://www.blender.org}}. The reader is referred to \cite{Kent13, Kent15} for a thorough description of the software in the context of astrophysical data visualization. 

{\blender} is implemented with a {\python} 3 application programming interface (API) which includes a built-in terminal and text windows for internal scripting, and the ability to write and execute external scripts without invoking the {\blender} graphical user interface (GUI). This {\python} API contains several major packages by default (e.g. \textsc{numpy}) and the ability to install additional external packages. For the large (and growing) fraction of astronomers already familiar with {\python}, {\blender} can be thought of as a new external visualization package. {\blender} also supports the {\xd} file format by default. 

\begin{figure}[htb!]
\centerline{\includegraphics[width=1.0\columnwidth]{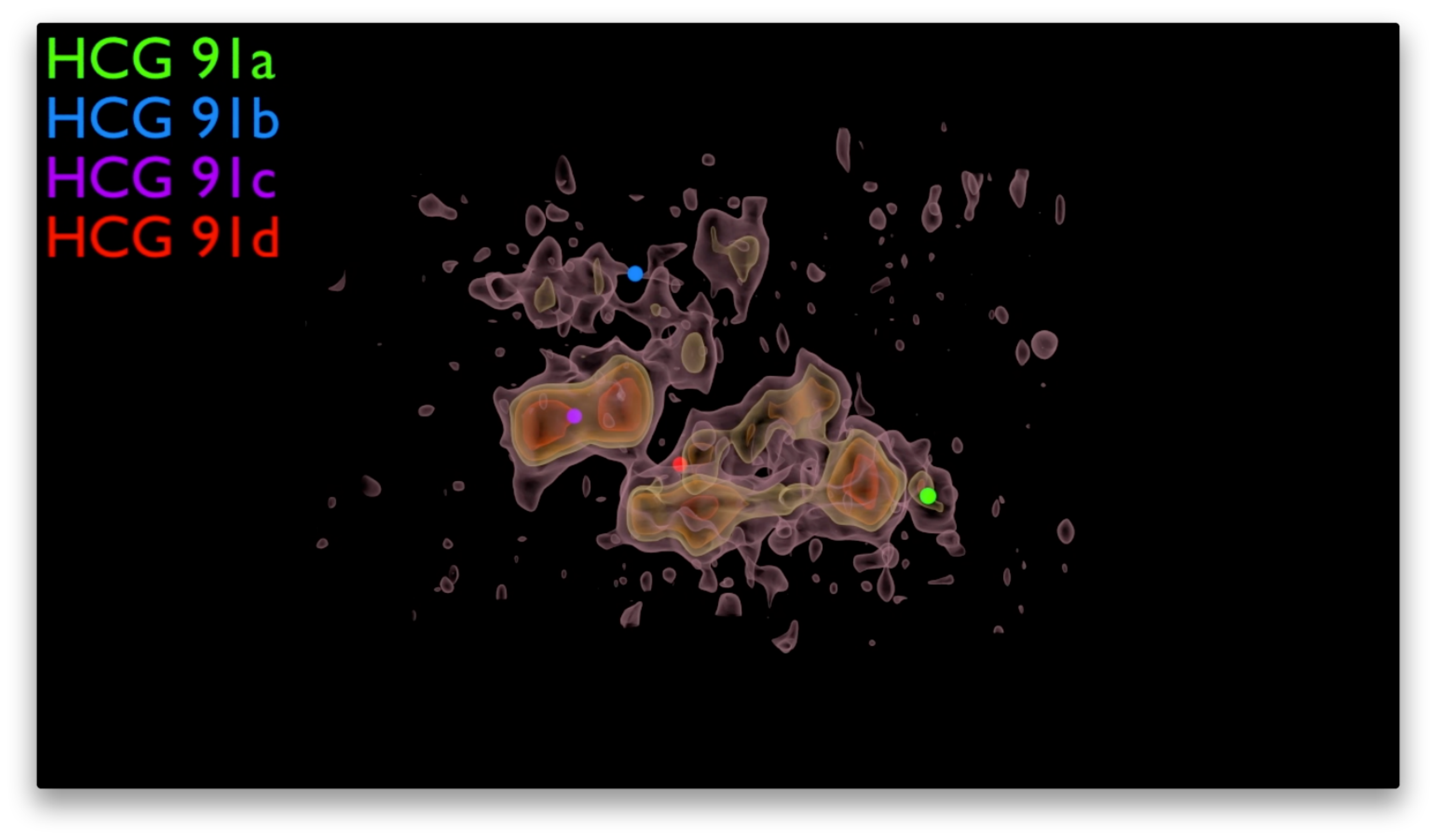}}
\caption{Final {\blender} rendered view of the {\HI} iso-intensity contours in HCG~91 observed by the \textit{VLA} (first exported to {\xd} using {\mayavi}) generated using the {\python} script included in the Github repository associated to this article. The corresponding movie is directly accessible via the DOI: \href{https://dx.doi.org/TBD}{TBD}. \textit{For arXiv readers: the movie is available here until publication: \href{http://www.mso.anu.edu.au/\~fvogt/website/arXiv/x3d/HCG91.mov}{http://www.mso.anu.edu.au/$\sim$fvogt/website/arXiv/ x3d/HCG91.mov}}.}\label{fig:blender}
\end{figure}

As a dedicated open-source 3-D creation suite software supporting ``\textit{modelling, rigging, animation, simulation, rendering, compositing, motion tracking, video editing and game creation}''\footnote{ \url{http://www.blender.org/about/}}, the number of options available for presentation and animation of {\xd} meshes in {\blender} is extremely large. The many capabilities of this software lie at the core of its steep learning curve, so that using {\blender} for the first time may appear as a formidable task. To ease the process and facilitate the first-time use of {\blender}, we provide in the Github repository associated to this article annotated {\python} scripts which automatically import {\xd} files generated using {\mayavi} (see Section~\ref{sec:python}) and produce high quality ready-to-render animations. One final rendered movie frame is presented in Figure~\ref{fig:blender} for completeness.

We note that our example scripts only explore a very limited number of features of {\blender}. We also include in the Github repository associated with this article a second animation of the \textit{VLA} datacube generated using a more complex version of the \textit{HCG91\_animation.py} script (available on demand) to illustrate some additional features of {\blender}. For more extensive examples and tutorials on how to use {\blender} with astrophysical datasets (including different ways to import the data and by-passing the {\xd} file format altogether), we refer the interested reader to the \textsc{astroblend} project \footnote{ \url{http://www.astroblend.com}}.

\section{Conclusions}\label{sec:conclusions}

Most datasets in modern Astrophysics are multi-dimensional. Yet, despite a largely electronic publishing landscape, their visualization and publication in dedicated articles remains problematic. Here, we have introduced the concept of the {\xd} pathway as a new approach to the creation, visualization and publication of 3-D diagrams in scientific articles. The fact that the {\xd} open-source (and ISO standard) 3-D file format lies at the center of an advanced data visualization product tree that includes interactive {\html} documents, 3-D printing, high-end animations (and possibly interactive {\pdf} documents) forms the core of the {\xd} pathway. 

The {\xd} pathway is designed to be ``file format-dependant'' rather than ``software-dependant'', so that it is less prone to rapid and un-anticipated changes and evolution. In particular, although the {\xd} standard is not immune to any evolution, such evolution will always be slower than that of associated software. Backwards compatibility is also a core characteristic of the {\xd} standard. We note that the evolution of the {\xd} standard can be directly influenced by official working groups. As such, the hypothetical creation of a dedicated \textit{Astronomy and Astrophysics Web3D working group} to promote and develop new {\xd} specifications for our field appears as an enticing idea.

The interactive {\html} branch of the {\xd} pathway is actively supported by leading peer-reviewed journals within the field of Astrophysics. From this perspective, we believe that the {\xd} pathway has already evolved beyond an ``experimental'' status. The future of the 3-D printing branch remains to be defined at this stage, but the possibility of rapidly creating 3-D printable models from existing {\xd} files ought to allow and facilitate additional experimentation with this technology. Finally, the {\xd} file format also provides ``regular'' astronomers the means to implement scientifically accurate and visually compelling movies and animations to communicate the complex, multi-dimensional content of their datasets. 

It has become increasingly evident that the large volume of modern multi-dimensional astrophysical datasets - either theoretical simulations or data products from existing/forthcoming observing facilities (e.g. the \textit{SKA}) - require innovative alternatives to 2-D diagrams for their visualization and publication. We strongly believe that tackling this issue requires to move the focus away from specific software, and direct the discussion towards file formats. Here, we propose that the {\xd} pathway offers a viable framework to develop alternative visualization techniques (such as interactive {\html} documents) while providing a single entry point connecting a growing range of advanced data visualization solutions. 

The detailed examples included in the Github repository associated to this article demonstrate the current possibilities (and existing limitations) offered by the {\xd} pathway in terms of data visualization, and provide the interested readers with the means of testing and implementing the {\xd} pathway with their own datasets. As a complete and new framework for multi-dimensional data visualization, we believe that the {\xd} pathway also provides a useful reference and comparison point (in terms of capabilities, limitations and flexibility) for ongoing efforts and future propositions regarding the visualization and publication of complex multi-dimensional datasets. 

\acknowledgments
We thank Colin Vest for performing the 3-D prints described in Section~\ref{sec:3dprinting}, Antoine Goutenoir for stimulating discussions and his feedback on some of the example scripts included in the Github repository associated to this article, and the anonymous referee for his/her comments and suggestions. The green and red dice examples presented in this article were first generated by Vogt as part of a lecture on advanced data visualization at the Australian National Institute for Theoretical Astrophysics (ANITA) Astroinformatics Summer School held in February 2015 in Canberra, and included in his Ph.D. thesis at the Australian National University (DOI:\href{http://dx.doi.org/10.4225/13/5553E63D6A79A}{10.4225/13/5553E63D6A79A}). This work has been partly supported by Grant AYA2011-30491-C02-01, and AYA2014-52013-C2-1-R, co-financed by MICINN and FEDER funds, and the Junta de Andalusia (Spain) Grant TIC-114. This research has made use of NASAÕs Astrophysics Data System and the NASA/IPAC Extragalactic Database (NED) which is operated by the Jet Propulsion Laboratory, California Institute of Technology, under contract with the National Aeronautics and Space Administration. This research has also made use of \textsc{astropy}, a community-developed core \textsc{python} package for Astronomy \citep{astropy13}, of \textsc{mayavi} \citep{Ram11}, of \textsc{aplpy}, an open-source plotting package for \textsc{python} hosted at http://aplpy.github.com, and of \textsc{montage}, funded by the National Aeronautics and Space Administration's Earth Science Technology Office, Computation Technologies Project, under Cooperative Agreement Number NCC5-626 between NASA and the California Institute of Technology.

\bibliographystyle{apj}
\bibliography{Vogt_Owen_2015}

\appendix
\section{Structure \& content of the Github repository associated to this article}\label{app:supp}
We include in a dedicated Github repository (see \url{https://github.com/fpavogt/x3d-pathway}, and/or DOI:\href{http://dx.doi.org/10.5281/zenodo.31953}{10.5281/zenodo.31953}) a series of scripts, data files and instruction sets (released as free software under the GNU General Public License version 3) designed to provide a stepping stone for the reader interested in implementing the {\xd} pathway. As an example, some of these scripts aided in the creation of the interactive counterpart to Figure~3(c) in \cite{Bouche15}. 

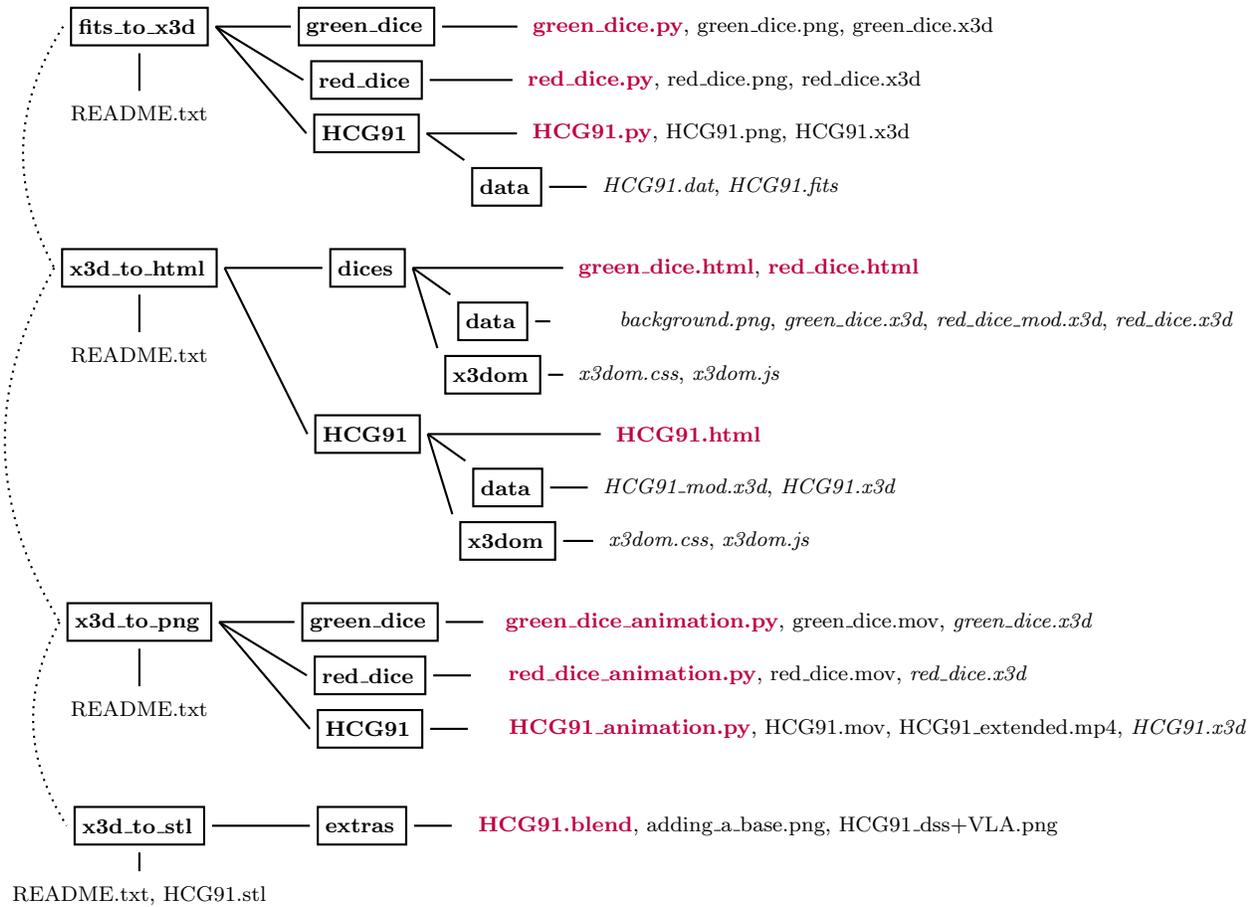
\begin{figure*}[htb!]
\begin{center}
\tikzstyle{every node}=[font=\footnotesize]
\tikzstyle{folder}=[rectangle,thick,minimum size=0.5cm,draw=black!100,fill=black!0,text centered,]
\tikzstyle{file}=[rectangle,thick,minimum size=0.5cm,draw=black!100,fill=black!5,text centered,]
\begin{tikzpicture}[auto, outer sep=3pt, node distance=5.5cm and 2cm,>=triangle 45]

\node[folder, text=black,] (x3d) {\textbf{fits\_to\_x3d\,}};
\node[folder, text=black, below=2.5cm of x3d] (html)  {\textbf{x3d\_to\_html\,}};
\node[folder, text=black, below=4.0cm of html] (png)  {\textbf{x3d\_to\_png\,}};
\node[folder, text=black, below=2.0cm of png] (stl)  {\textbf{x3d\_to\_stl\,}};

\node[folder, text=black, right = 1 cm of x3d] (x3d_green)  {\textbf{green\_dice\,}};
\node[folder, text=black, below=0.cm of x3d_green] (x3d_red) {\textbf{red\_dice\,}};
\node[folder, text=black, below =0. of x3d_red] (x3d_HI)  {\textbf{HCG91\,}};
\node[folder, text=black, below right=0. and 0.5cm of x3d_HI] (x3d_HI_data)  {\textbf{data\,}};

\node[folder, text=black, right = 1.3 of html] (html_dices)  {\textbf{dices\,}};
\node[folder, text=black, below right=0. and 0.5cm of html_dices] (html_dices_data)  {\textbf{data\,}};
\node[folder, text=black, below=0. of html_dices_data] (html_dices_x3dom)  {\textbf{x3dom\,}};
\node[folder, text=black, below =1.5 of html_dices] (html_HI)  {\textbf{HCG91\,}};
\node[folder, text=black, below right=0. and 0.5cm of html_HI] (html_HI_data)  {\textbf{data\,}};
\node[folder, text=black, below=0. of html_HI_data] (html_HI_x3dom)  {\textbf{x3dom\,}};

\node[folder, text=black, right = 1 cm of png] (png_green)  {\textbf{green\_dice\,}};
\node[folder, text=black, below=0. cm of png_green] (png_red) {\textbf{red\_dice\,}};
\node[folder, text=black, below =0. of png_red] (png_HI)  {\textbf{HCG91\,}};

\node[folder, text=black, right= 1.3 of stl] (stl_extra) {\textbf{extras\,}};


\node[below = .5 cm of x3d] (x3d_readme) {README.txt};
\node[below = .5 cm of html] (html_readme) {README.txt};
\node[below = .5 cm of png] (png_readme) {README.txt};
\node[below = .25 cm of stl] (stl_files) {{README.txt}, {\color{black} HCG91.stl}};

\node[right = 1.cm of x3d_green] (x3d_green_files) {\textbf{\color{purple} green\_dice.py}, {\color{black} green\_dice.png}, {\color{black} green\_dice.x3d} }; 
\node[right=1.1cm of x3d_red] (x3d_red_files) {\textbf{\color{purple} red\_dice.py}, {\color{black} red\_dice.png}, {\color{black} red\_dice.x3d} }; 
\node[right=1.2cm of x3d_HI] (x3d_HI_files) {\textbf{\color{purple} HCG91.py}, {\color{black} HCG91.png}, {\color{black} HCG91.x3d} }; 
\node[right=0.5 of x3d_HI_data] (x3d_HI_data_files) {\textit{HCG91.dat}, \textit{HCG91.fits} }; 

\node[right =2.cm of html_dices] (html_dices_files) {\textbf{\color{purple} green\_dice.html}, \textbf{\color{purple} red\_dice.html}};
\node[right =0.2 of html_dices_data, text width=9.6cm, align=center] (html_dices_data_files) {\textit{background.png}, \textit{green\_dice.x3d}, \textit{red\_dice\_mod.x3d}, \textit{red\_dice.x3d}};
\node[right = 0.2 of html_dices_x3dom] (html_dices_x3dom_files) {\textit{x3dom.css}, \textit{x3dom.js}};

\node[right = 2.3cm of html_HI] (html_HI_files) {\textbf{\color{purple} HCG91.html}};
\node[right =.5 of html_HI_data] (html_HI_data_files) {\textit{HCG91\_mod.x3d}, \textit{HCG91.x3d}};
\node[right = .4cm of html_HI_x3dom] (html_HI_x3dom_files) {\textit{x3dom.css}, \textit{x3dom.js}};

\node[right =0.5 of png_green, text width=8.cm, align=center] (png_green_files) {\textbf{\color{purple}green\_dice\_animation.py}, {\color{black}green\_dice.mov}, \textit{green\_dice.x3d}};
\node[right = 0.5 of png_red, text width=7.5cm, align=center] (png_red_files) {\textbf{\color{purple}red\_dice\_animation.py}, {\color{black}red\_dice.mov}, \textit{red\_dice.x3d}};
\node[right = 0.5 of png_HI, text width=10.5cm, align=center] (png_HI_files) {\textbf{\color{purple}HCG91\_animation.py}, {\color{black}HCG91.mov}, HCG91\_extended.mp4, \textit{HCG91.x3d}};

\node[right = .5 of stl_extra, text width=8cm, align=center] (stl_extra_files) { \textbf{\color{purple} HCG91.blend}, {\color{black} adding\_a\_base.png}, {\color{black} HCG91\_dss+VLA.png}};


\draw[-, thick, black](x3d) to node {} (x3d_readme);
\draw[-, thick, black](html) to node {} (html_readme);
\draw[-, thick, black](png) to node {} (png_readme);

\draw[dotted,thick, black, bend right] (x3d.west) to node {} (html.west);
\draw[dotted,thick, black, bend right] (html.west) to node {} (png.west);
\draw[dotted,thick, black, bend right] (png.west) to node {} (stl.west);

\draw[-,thick, black] (x3d.east) to node {} (x3d_green.west);
\draw[-,thick, black] (x3d.east) to node {} (x3d_red.west);
\draw[-,thick, black] (x3d.east) to node {} (x3d_HI.west);

\draw[-,thick, black] (x3d_green) to node {} (x3d_green_files);
\draw[-,thick, black] (x3d_red) to node {} (x3d_red_files);
\draw[-,thick, black] (x3d_HI) to node {} (x3d_HI_files);
\draw[-,thick, black] (x3d_HI.east) to node {} (x3d_HI_data.north west);
\draw[-,thick, black] (x3d_HI_data) to node {} (x3d_HI_data_files);

\draw[-,thick, black] (html.east) to node {} (html_dices.west);
\draw[-,thick, black] (html_dices) to node {} (html_dices_files);
\draw[-,thick, black] (html_dices.east) to node {} (html_dices_data.north west);
\draw[-,thick, black] (html_dices.east) to node {} (html_dices_x3dom.north west);
\draw[-,thick, black] (html_dices_data) to node {} (html_dices_data_files);
\draw[-,thick, black] (html_dices_x3dom) to node {} (html_dices_x3dom_files);
\draw[-,thick, black] (html.east) to node {} (html_HI.west);
\draw[-,thick, black] (html_HI) to node {} (html_HI_files);
\draw[-,thick, black] (html_HI.east) to node {} (html_HI_data.north west);
\draw[-,thick, black] (html_HI.east) to node {} (html_HI_x3dom.north west);
\draw[-,thick, black] (html_HI_data) to node {} (html_HI_data_files);
\draw[-,thick, black] (html_HI_x3dom) to node {} (html_HI_x3dom_files);

\draw[-,thick, black] (png.east) to node {} (png_green.west);
\draw[-,thick, black] (png.east) to node {} (png_red.west);
\draw[-,thick, black] (png_green.east) to node {} (png_green_files);
\draw[-,thick, black] (png_red.east) to node {} (png_red_files);
\draw[-,thick,  black] (png.east) to node {} (png_HI.west);
\draw[-,thick, black] (png_HI.east) to node {} (png_HI_files);

\draw[-,thick, black] (stl) to node {} (stl_files);
\draw[-,thick, black] (stl.east) to node {} (stl_extra.west);
\draw[-,thick, black] (stl_extra) to node {} (stl_extra_files);

\end{tikzpicture}
\vspace{10pt}
\caption{Structure and content of the Github repository associated to this article (at the time of publication). Folders are marked with boxes. Scripts and instruction files are marked in bold purple. Required data input files are in italic. Four dedicated ``README'' files provide specific introduction to the different families of examples. This material can be downloaded as free software (under the GNU General Public License version 3) via the DOI:\href{http://dx.doi.org/10.5281/zenodo.31953}{10.5281/zenodo.31953}. }\label{fig:supp_mat}
\end{center}
\end{figure*}

A global overview of the structure and content of Github repository (at the time of publication of this article) is provided in Figure~\ref{fig:supp_mat}. The material is divided in four sections dedicated to the creation of {\xd} files, the creation of interactive {\html} documents, the creation of high-end animations and 3-D printing. The scripts described in this article are contained in the release v0.9 associated with this Github directory. All scripts provided are commented in detail to help in their understanding. 

The different examples build on one another: for example, the {\xd} files created using the {\python} scripts are used as input material for the {\html} visualization scripts. Included are three examples of incremental complexity. The ``green dice'' examples (\textit{green\_dice.py, green\_dice.html} and \textit{green\_dice\_animation.py}) are the most basic. The red dice examples were designed to hit some of the limitations associated with {\mayavi} to export {\xd} files. The third set of examples (\textit{HCG91.py, HCG91.html} and \textit{HCG91\_animation.py}) are the most complex, and do not shy away from the additional intricacies associated with visualizing real astrophysical datasets, including the use of \textit{World Coordinate System} units.

\end{document}